# Thermoelectric enhanced *ZT* regime calculated by Fermi integral method

Hirofumi Kakemoto[1]

[1]*Clean Energy Research Center, University of Yamanashi, 4-3-11, Takeda, Kofu, Yamanashi 400-8511, Japan*

We report about dimensionless figure of merit ($ZT$) calculated by using Fermi integral method (compared with $Bi_2Te_3$, $CoSb_3$, $SrTiO_3$) for thermoelectric (TE) materials' design and its module application. Particularly, TE properties: electrical conductivity (small polaron: $\sigma(m^*)$), Seebeck coefficient ($S$), thermal conductivity ($\kappa_e$), and $ZT$ were calculated by using reduced energy ($\zeta=E/k_BT$), as the functions of $T$, and effective mass ($m^*/m$). Enhancement of $ZT$ was investigated by the contour plots of $T$, and $m^*/m$ versus $\zeta$.

## 1. Introduction

Nowadays, renewed and recycled energies from sun light heat or waste heat are demanded to realize the sustainable society. Thermoelectric (TE) material is possible to generate thermal electromotive force (EMF) from several waste heats. Recently, TE material, such as metallic: $Bi_2Te_3$,[1] $CoSb_3$ (p-type),[2] SiGe ($n$-type), and oxide: $Na_xCoO_2$ ($p$-type), and $SrTiO_3$ [3] ($n$-type) show high TE properties. Particularly, more high-performance $n$-type TE oxide material is demanded to develop for realizing high performance $p$-$n$ pairs of TE module use.

To design TE material, 1) Fermi integral [4-9] or Boltzmann equation by using band structure ($E_g$ and $E_F$ etc) [10], 2) Heikes formula [11], and 3) space charge [12] theories are reported. Electrical conductivity ($\sigma$), Seebeck coefficient ($S$), and thermal conductivity ($\kappa_e=L\sigma T$) are able to be expressed by carrier density ($N$), finally expressed by Fermi integral ($F_r$), hence design for enhancement of TE properties is not simple. Enhancement of dimensionless figure of merit ($ZT$) is reported about i) small polaron or ii) rattling, with increasing $S$ and decreasing $k_{ph}$.

Hicks and Dresselhaus presented the low dimensional density of state model and $B$-parameter for increasing $ZT$ by using Fermi integral method. [1]

Mahan also reported about $Z$ calculation. The highest $Z$ is calculated from near band edge, and $Z$ is close to $B$ -parameter. [7,8]

In addition, small polaron conductivity ($\sigma(m^*)$) and its band structure model are reported for many TE materials. [11] Here, $ZT$ is defined as $S^2\sigma T/(\kappa_e+\kappa_{ph})$, and $ZT$ can be modified as $S^2/[L+\kappa_{ph}/\sigma(m^*)T]$ by using $\sigma(m^*)$. $\sigma(m^*)$ is increased with increasing $\zeta_F$. Therefore, $\sigma(m^*)$ gives a direction for enhancement of $ZT$.

We report about enhancement of $ZT$ about TE material calculated by using Fermi integral method as the functions of reduced energy ($\zeta_F$) and effective mass ($m^*$).

## 2. Calculation

$\sigma$, $S$ and $\kappa_e$ ($=L\sigma T$) were calculated using eq.(1)–(4) by using Fermi integral method. Fermi integral ($F_r(\zeta_F)$) is defined as $\int \xi^r d\xi/[\exp(\xi-\zeta_F)+1]$, where $\zeta_F$ ($\geqq -0.05$ for $\sigma$ and $S$), and $\tau$ are $E_F/k_BT$ ($T$=300K-1000K) and $\tau_0\varepsilon^{r-1/2}$, respectively, and Fermi integral can be approximated as $F_r(\zeta_F)=\Gamma(r+1)e^{\zeta F}\,\Sigma(-1)^j e^{j\zeta F}/(j+1)^{r+1}$, while $\zeta < -0.05$ for $\sigma$, $S$, and $n>-1$.[9]

$$N=(1/2\pi^2)(2k_BT/\hbar^2)^{3/2}m^{*3/2}F_{1/2}(\zeta_F), \qquad (1)$$

$$\sigma(m^*)=(1/2\pi^2)(2k_BT/\hbar^2)^{3/2}m^{*3/2}F_{1/2}(\zeta_F)q\mu, \qquad (2)$$

$$S=-(k_B/q)\{\delta(\zeta_F)-\zeta_F\},\ \delta(\zeta_F)=5F_{3/2}(\zeta_F)/3F_{1/2}(\zeta_F), \qquad (3)$$

$$L=(2k_B^2\tau/3qm^{*1/3}\mu)(7F_{2+1/2}/2F_{1/2}-25F^2_{1+1/2}/6F^2_{1/2}). \qquad (4)$$

Dimensionless figure of merit ($ZT$) is expressed as

$$ZT=S^2\sigma T/(\kappa_e+\kappa_{ph})=S^2/[L+\kappa_{ph}/\sigma(m^*)T]. \qquad (5)$$

## 3. Results and discussion

Figure 1 (a) shows reported $Z$ for $n$-type $Bi_2Te_3$, $n$-type $Si_{1-x}Ge_x$, $p$-type $CoSb_3$ and $n$-type $SrTiO_3$ (dots and solid lines), and $ZT$ (solid lines: 0.1, 1 and 2) as a function of $T$. $Bi_2Te_3$ reported up to $ZT$>1 at 400K, $CoSb_3$ takes $ZT$~0.1 at 600K, $Si_{1-x}Ge_x$ shows $ZT$~1 at around 1200K, and $SrTiO_3$ takes $ZT$ ~0.8 at 1000K. [13]

The classical $S$, $\sigma$ versus logarithmic carrier density (log$N$) (Ioffe phenomena) is plotted in Fig.1 (b). $\sigma$ is possible to represent the enhancement by increased $m^*$: small polaron effect, but power factor ($S^2\sigma$) term of $ZT$ cannot be increased from the simple relationship between $S$ and $\sigma(m^*)$ with $m^*/m$ =1.

The mobility ($\mu$) of TE materials versus $m^*/m$ is plotted in Fig.1 (c). Small polaron is reported in $\mu$ of $Bi_2Te_3$: 1200 $cm^2/Vs$ at 300K[1], $CoSb_3$: 310 $cm^2/Vs$ at 600K [2], and $SrTiO_3$: 0.5 $cm^2/Vs$ at 1000K. [3]

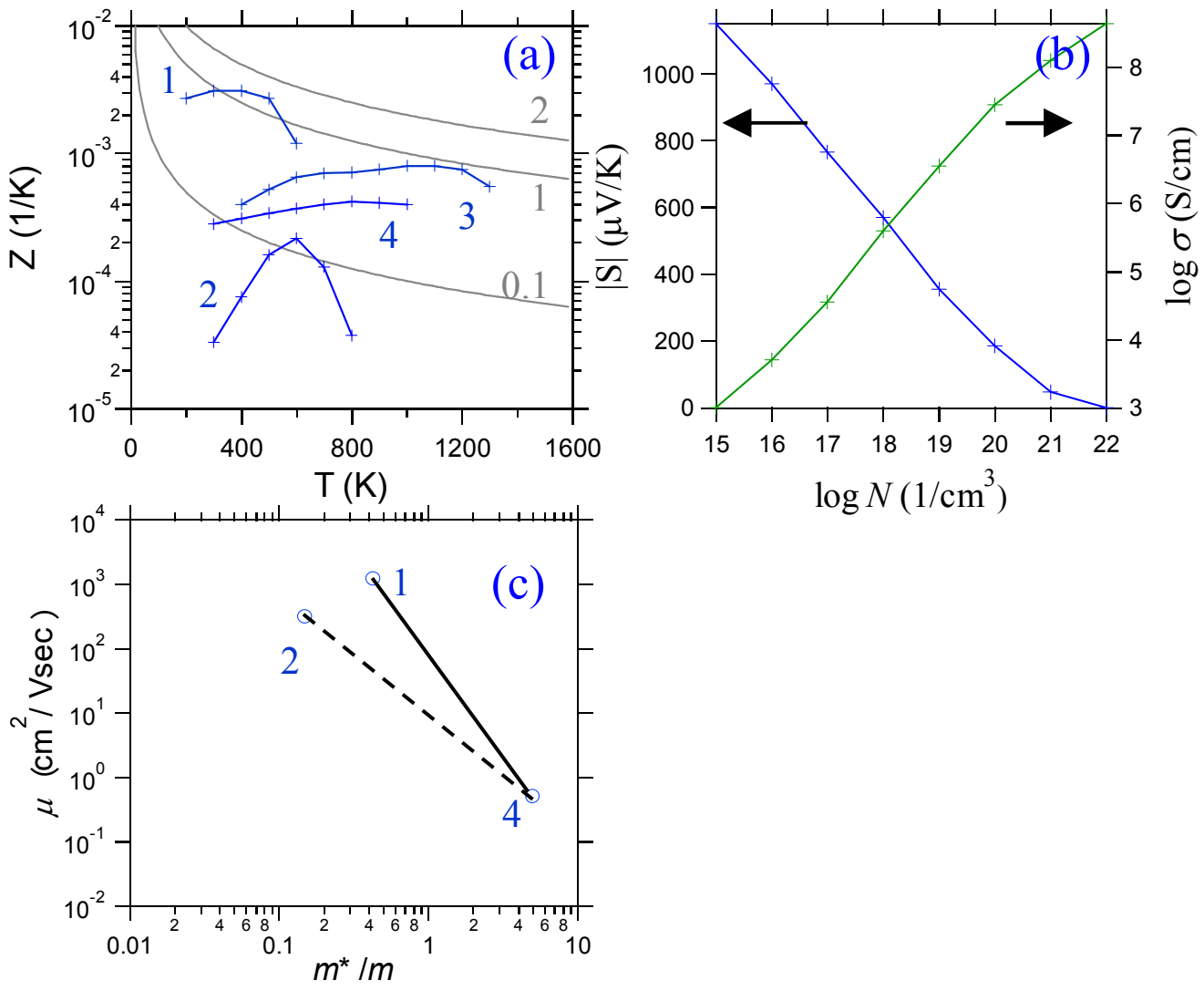

FIG.1 (a) $Z$ of $Bi_2Te_3$: 1, $CoSb_3$: 2, $Si_{1-x}Ge_x$: 3, and $SrTiO_3$: 4 (dots and solid lines), and $ZT$ (solid lines) versus $T$, (b) $S$, $\sigma$ versus log$N$ plot, (c) mobility ($\mu$) versus effective mass: $m^*/m$ (dots), and extrapolated line (solid and dashed lines) [14].

With comparing these experimental values, we try to reproduce theoretically these values by using Fermi integral method. The calculating conditions for $F_r$: $m^*/m$, $\xi$, and $T$ are listed in Table I. In calculation, $E_F$ was selected from −66.4meV to 0.01eV, at $T$=300K~1000K, and was input to $\zeta_F=E_F/k_BT$ in eq.(1)~(4).

Here, we introduce $m^*$ for $\sigma$, and $L$ in eq.(2),(4): ($Bi_2Te_3$, $CoSb_3$, $SrTiO_3$: $m^*/m$=0.43, 0.15, 5.0, respectively) as listed in Table I. In addition, Fermi integral and/or its approximation are selected with comparing results of calculation.

Here, $\xi$ in $F_r$ is chosen from −5 to 5. By using eq.(1-4), $\sigma$, $S$ and $L$ were calculated for $n$-type $Bi_2Te_3$ (condition I), $p$-type $CoSb_3$ (condition II), and $n$-type $SrTiO_3$ (condition III) as also shown in Table I. Figure 2 plots results of calculation in TE properties ($\sigma$, $S$, $L$) by using Fermi integral method. $\sigma$ varies from $10^{-2}$ to $10^4$ S/cm, $S$ shows from +500 $\mu$V/K: $p$-type to −500 $\mu$V/K: $n$-type, and $L$ takes from $10^{-5}$ to $10^{-2}$ W$\Omega$/K$^2$ as a function of $\zeta_F$ (>0).

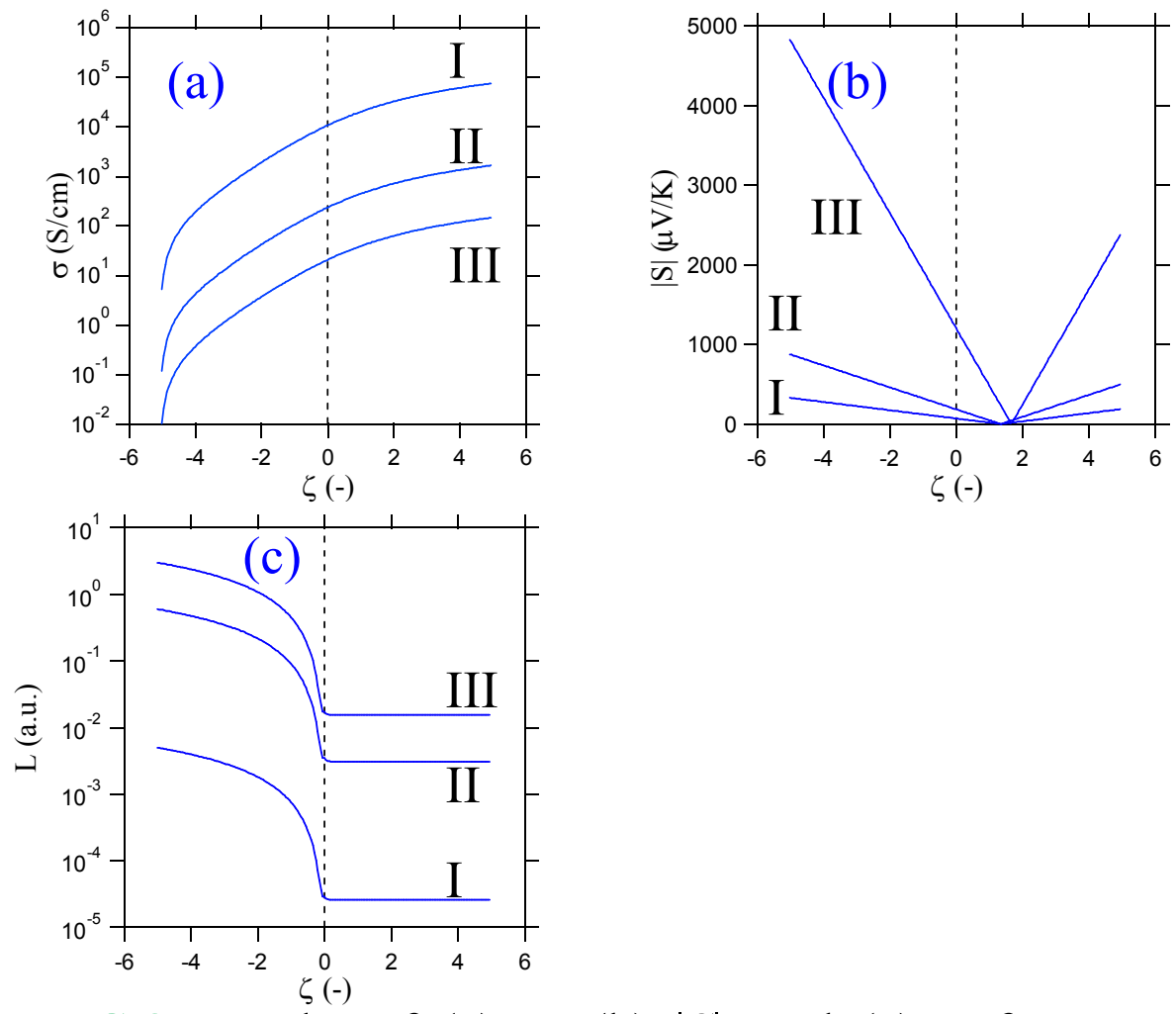

FIG.2 Results of (a) $\sigma$, (b) |$S$|, and (c) $L$ for $n$-type $Bi_2Te_3$ (condition I), $p$-type $CoSb_3$ (condition II), and $n$-type $SrTiO_3$ (condition III) calculated by using Fermi integral method introducing $m^*$ in Table I.

From Fig.2, $ZT$ is reproduced to be 1~2, ~0.1, 0.8~1.0 using eq.(5) for $Bi_2Te_3$, $CoSb_3$, and $SrTiO_3$ at $\zeta_F$= −0.05~+0.2.

The theoretical calculation for $Bi_2Te_3$ by using Fermi integral method was reported as $ZT$=0.52 at 300K (experimental data: $ZT$=0.67). Our result for $Bi_2Te_3$ in Fig.2 is close to above values.[1] The results of experimental data for $CoSb_3$ was reported as $ZT$=0.1 at 600K[2], and theoretical calculation by using Boltzmann eq. was also carried out.[10] Our result for $CoSb_3$ is consisted with above value.

The experimental study for $SrTiO_3$ was reported as $ZT$=0.1 at 1000K.[3] Our result for $SrTiO_3$ is consisted with above value.

Table I Calculating conditions: band structure, carrier type, $m^*/m$, $\xi$, and $T$.

| Material | $E_c$, $E_v$, $E_F$, $E_g$ (eV) | type | $m^*/m$ | $\xi$ (-) | $T$(K) | Condition | Ref. |
|---|---|---|---|---|---|---|---|
| PbTe | | _ | 0.05 | | 500K | | |
| $Bi_2Te_3$ | 0.05, −0.05, 0.01, 0.13 | $n$ | 0.43 | −5~5 | 350~400K | I | [1] |
| $CoSb_3$ | 0.1, −0.1, ±66.4meV, 0.63 | $p$ | 0.15 | −5~5 | 550~600K | II | [2] |
| $Si_{1-x}Ge_x$ | | $n$ | 1.06 | | 1100K | | |
| $SrTiO_3$ | 1.5, -1.5, ~0, 3 | $n$ | 5.00 | −5~5 | 950~1000K | III | [3] |

Figure 3 shows contour plots of $ZT$, $T$, and $m^*/m$ versus $\zeta_F$ in the range from −5 to 3. In Figs.3(a, b), the results of $Bi_2Te_3$, from condition I: (by using eq.(3-5), $n$-type, $S= -60\mu V/K$, $\sigma=1.1\times10^4$ S/cm, $\kappa=1.5$ W/mK, at $0.42<m^*/m<0.44$, 350K< $T$ <400K, and $\zeta_F= +0.2$) are estimated, and $ZT$ is possible to be enhanced up to 1.1. In Figs.3(c, d), the results of $CoSb_3$, from condition II: (by using eq.(3-5), $p$-type, $S=+192$ $\mu$V/K, $\sigma=250$ S/cm, $\kappa=4.8$ W/mK, at $0.14<m^*/m<0.16$, 550K< $T$ <600K, and $\zeta_F= -0.05$), $ZT$ is also slightly enhanced up to 0.1. In Figs.3(e, f), the results of $SrTiO_3$, condition III: (by using eq.(3-5), $|S|=1243\mu V/K$, $\sigma=67$S/cm, $\kappa=10$W/mK, at $4.9<m^*/m<5.1$, 950K< $T$ <1000K, and $|\zeta_F|=0.05$). $ZT$ is also estimated to be 0.9. From calculation results, $ZT$ can be reproduced the enhancement by small polaron conductivity ($\sigma(m^*)$.

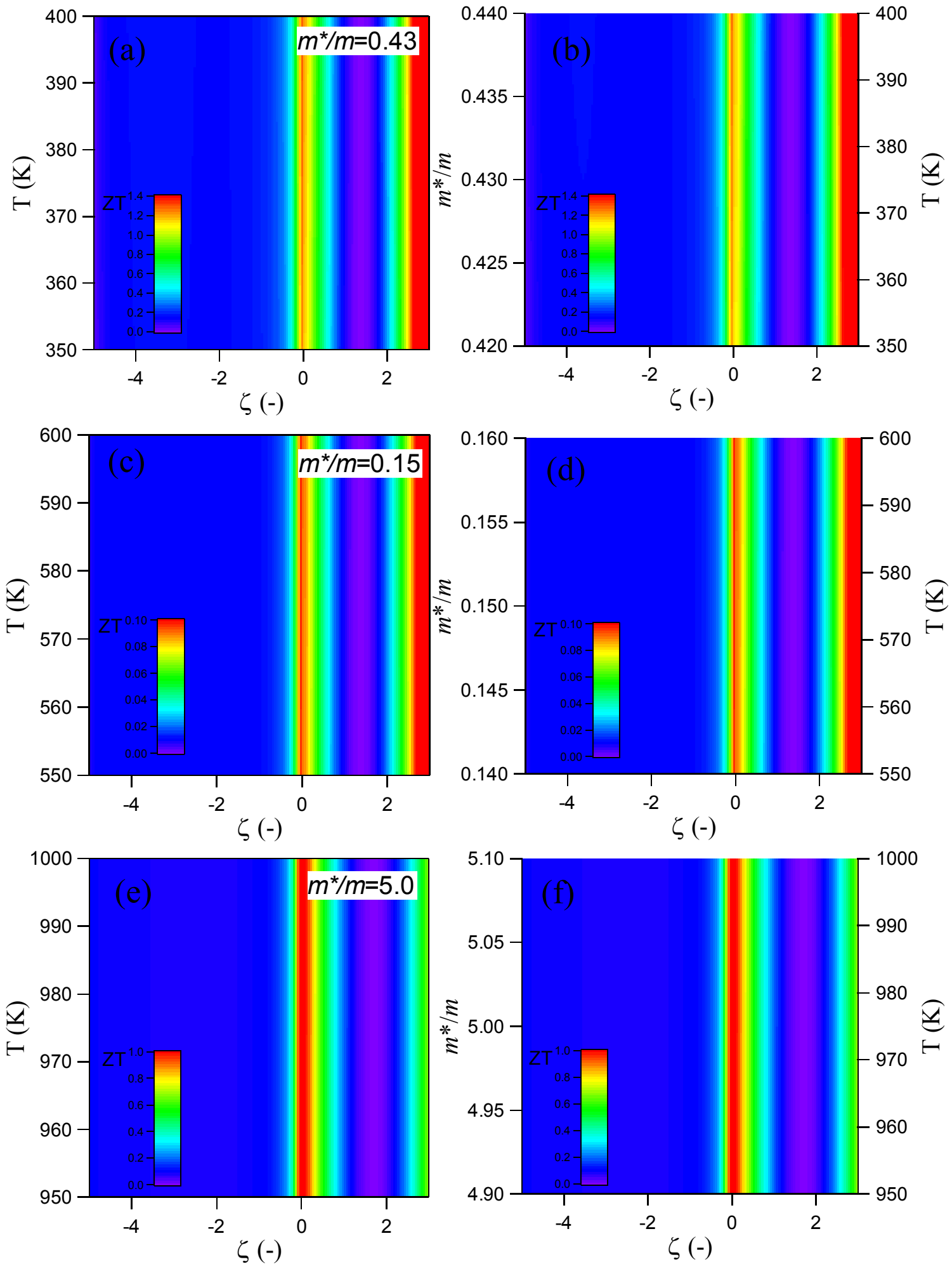

FIG. 3 $ZT$ contour plots of $T$, and $m^*/m$ versus $\zeta_F$ calculated by using Fermi integral method, (a,b) $Bi_2Te_3$, condition I (at $m^*/m$: 0.42~0.44, and 350~400K), (c,d) $CoSb_3$, condition II (at $m^*/m$: 0.14~0.16, and 550~600K), and (e,f) $SrTiO_3$, condition III (at $m^*/m$: 4.9~5.1, and 950~1000K).

## 4. Conclusion

In this report, electrical conductivity affected by small polaron: $\sigma(m^*)$ was introduced to calculate the dimensionless figure of merit ($ZT$) for thermoelectric materials: $n$-type $Bi_2Te_3$, $p$-type $CoSb_3$, and $n$-type $SrTiO_3$, and Seebeck coefficient ($S$), thermal conductivity ($\kappa_e$), and $ZT$ were carried out the calculation by using Fermi integral method.

Enhanced $ZT$ were estimated its contour plots of $T$ and $m^*/m$ versus reduced energy ($\zeta_F$).

In future study, the investigation of $n$-type Nb related TE oxide will be carried out using Fermi integral method.

Acknowledgment

This work was partly supported by Japan Society for the Promotion of Science (JSPS) KAKENHI Grant-in-Aid for Scientific Research(C) Number JP25410238.

*Present address: 1-15-11, Sakura-cho, Tsuchiura, Ibaraki, 300-0037, Japan, Techno Pro R&D company (Tsukuba branch), Techno Pro Inc.

e-mail: hkakemoto@yamanashi.ac.jp